\documentclass[letters,fleqn,usenatbib]{mnras}

\usepackage{newtxtext,newtxmath}

\usepackage{float}

\usepackage[T1]{fontenc}
\usepackage{ae,aecompl}

\usepackage{graphicx}	
\usepackage{amsmath}	
\usepackage{amssymb}	
\usepackage[colorinlistoftodos]{todonotes}
\usepackage{helvet}
\usepackage{color,colortbl}

\title[CVs in {\it Gaia's} HRD]{Disentangling Cataclysmic Variables in {\it Gaia's} HR-Diagram}

\author[J. Abril et al.]{
Javier Abril,$^{1,3}$\thanks{E-mail: Jabrilib@eso.org}
Linda Schmidtobreick,$^{1}$ 
Alessandro Ederoclite,$^{2}$ \newauthor
and Carlos L\'opez-Sanjuan$^{3}$
\\
$^{1}$European Southern Observatory (ESO), Alonso de C\'ordova 3107, Vitacura, Santiago, Chile\\
$^{2}$Instituto de Astronomia, Geof\'isica e Ci\~encias Atmosf\'ericas (IAG), Universidade de S\~ao Paulo (USP), Rua do Mat\~ao 1226, \\C. Universit\'aria, S\~ao Paulo, 05508-090, Brazil\\
$^{3}$Centro de Estudios de F\'isica del Cosmos de Arag\'on (CEFCA) - Unidad Asociada al CSIC, Plaza San Juan, 1, E-44001, Teruel, Spain
}

\date{Accepted XXX. Received YYY; in original form ZZZ}

\pubyear{2019}

\begin{document}
\label{firstpage}
\pagerange{\pageref{firstpage}--\pageref{lastpage}}
\maketitle

\begin{abstract}
Cataclysmic Variables (CVs) are interacting binaries consisting of at least three components that control their colour and magnitude. Using {\it Gaia} we here investigate the influence of the physical properties of these binaries on their position in the Hertzsprung-Russell diagram (HR-diagram). The CVs are on average located between the main sequence and the white dwarf regime, the maximum density being at $G_{BP}-G_{RP} \sim 0.56$ and $G_{abs} \sim 10.15$. We find a trend of the orbital period with colour and absolute brightness: with decreasing period, the CVs become bluer and fainter. We also identify the location of the various CV sub-types in the HR-diagram and discuss the possible location of detached CVs, going through the orbital period gap.
\end{abstract}





\section{Introduction}

Cataclysmic Variables are semi-detached binaries built of a white dwarf (WD) which is accreting mass from a Roche-lobe filling main sequence (MS) star. Mass transfer is driven by the loss of angular momentum in absence of strong magnetic fields, and the transferred material forms an accretion disc surrounding the central WD (see e.g., \citealt{warner95-1}, \citealt{hellier_book} and \citealt{kniggeetal11-1} for comprehensive reviews). Since the structure of both components is relatively simple, CVs are one of the best sources to test our understanding of many astrophysical phenomena involving evolution of compact, interacting binaries and accretion phenomena. Their study helps to resolve standing discrepancies between current population models and observations in many present and complex topics including black hole binaries, short gamma-ray bursts, X-ray transients, milli-second pulsars and Supernovae Ia. 

The orbital period distribution is the main tool to study the evolution of CVs, as it presents features in key points that allow us to understand their behaviour. As a consequence of the angular momentum loss and the mechanisms driving it, CVs move from long orbital periods and high mass transfer rates to short orbital periods and low mass transfer rates \citealt{paczynski+sienkiewicz83-1}; \citealt{Townsley09}; \citealt{GoliaschNelson15}; \citealt{palaetal7}). The evolution proceeds in this way until the system reaches the ``period minimum'' at $\sim$ 76-80 minutes (\citealt{knigge06-1}; \citealt{gaensickeetal09}) in which the donor turns into a brown dwarf. Consequently, the orbital separation and period now increases as the mass transfer continues, becoming in the so-called period bouncers, faint systems with short orbital periods. On their way to the period minimum, observations show an abrupt drop in the number of systems with periods between 2 and 3h, referred to as the period gap. Below this range (Porb $<$ 2\,h) systems have low mass-transfer rates governed by gravitational radiation (GR) \citep{patterson84-1} while the higher mass-transfer rates above the gap (Porb $>$ 3\,h) are a consequence of the stronger magnetic braking (MB) (\citealt{rappaportetal83-1}; \citealt{spruit+ritter83-1}; \citealt{hameuryetal88-2}; \citealt{davis08}). The standard explanation suggests that MB switches off when a CV has evolved down to 3h, the secondary contracts to its thermal equilibrium and detaches from its Roche lobe. Such systems crossing the gap are known as detached Cataclysmic Variables (dCVs). The continuing angular momentum loss by GR shrinks the orbit until at a period of about 2h, the Roche lobe makes contact with the stellar surface again and mass transfer is re-established albeit at a lower level. 

During this evolution, the CV will change appearance: The relative contribution of the WD, the secondary star and the accretion disc or stream makes for unique colours \citep[e.g.][]{szkodyetal02-2}. The systems thus occupy distinct locations in colour-colour diagrams with respect to single stars. Due to the relatively small sample of CVs and the inherent difficulty of any source in obtaining its distance, it has not been possible so far to perform an analysis of the CV absolute magnitude distribution. 
With the arrival of {\it Gaia},
this has changed. Already \cite{Pala19} show the advances that {\it Gaia} parallaxes bring to the understanding of CVs, and we now have the data to study CVs in the HR-diagram.

The paper is organised as such: In section \ref{sec:gaia}, we explain how we use {\it Gaia} to define our CV sample. In section \ref{sec:HRdiagram} the results are presented for all CVs and the trends with the orbital periods and the subtypes are discussed. Finally, we present our summary in \ref{sec:Summary}.

\section{The Catalogue of {\it Gaia} DR2 and the cross-match with CV catalogue}\label{sec:gaia}

The goal of the {\it Gaia} 
space mission is to make  the largest, most precise three-dimensional map of the Milky Way to-date by detecting and measuring the motion and parallax of each star in its orbit around the centre of the Galaxy. To this means, the three filters $G$, $G_{BP}$ and $G_{RP}$ are observed at several epochs over a period of about 670 days of mission operations. For details, see \cite{GaiaDR2}.

The second data release (GDR2 hereafter) is based on 22 months of observations and provides positions, parallaxes and proper motions for 1.3 billion sources up to G $\sim$ 20 magnitudes. This kind of data allows the derivation of distances and absolute magnitudes to study the position of all objects in the global HR-diagram.

\subsection{Deriving absolute magnitudes}\label{subsubsec:Abs_mag}

One of the aims of this paper is finding the CV locus in the HR-diagram. We make use of GDR2 data to compute their absolute magnitudes. 
The absolute magnitude $M$ of an object is given by:
\begin{equation}
M = m + 5 - 5\log(d) + A,
\label{eq:M}
\end{equation}

where $m$ is the apparent magnitude, $A$ is the interstellar extinction and $d$, the distance to the source which can be obtained by the GDR2 data. 
GDR2 provides weighted mean fluxes\footnote{Weighted means are used because flux errors on different epochs may vary depending on the configuration of each observation, see \cite{CarrascoGDR1} and \cite{RielloGDR2} for detailed information.} and, as CVs are variable stars, this procedure has an effect on their $G$
, $G_{BP}$ and $G_{RP}$ values. The degree of impact might be determined by comparing the 670-days length of the {\it Gaia}-mission and the cycle of variation length for every CV subtype. The highest impact is on Dwarf Novae systems as they can have outbursts even on a weekly base (\citealt{SterkenBook} and references therein). The low recurrence in Novae, Nova-like and Magnetic CVs should have no significant impact on the overall sample.

Inferring the distance from the {\it Gaia} DR2 parallax is not a trivial issue. Distance can be derived as the inverse of the parallax, only if the parallax error is lower than 20\% and by doing so we would discard 80\% of the sources (\citealt{Luri18}). We used instead the distances inferred by \citet{Bailer-Jones18} who compute distances and their uncertainties through a probabilistic analysis based on the Bayes theorem and adopting an exponentially decreasing space density 
prior\footnote{For a detailed explanation of this approach and an analysis of applying this technique refer to \citet{Bailer-Jones18}. For a discussion of the use of different priors, see
\citet{Bailer-Jones18},\citet{Luri18},\citet{igoshevetal2016},\citet{astraatmadjaetal16}.
} 
for the 1.33 billion sources from GDR2. These are available using ADQL\footnote{http://gaia.ari.uni-heidelberg.de/tap.html} and we here use them to derive  the absolute magnitudes through Equation~\ref{eq:M}.

\subsection{The CV sample}\label{sec:CVsample}


The Catalog and Atlas of Cataclysmic Variables \citep{DownesCat} includes all objects which have been classified as a CV at some point in time. Although it was frozen on February 1st, 2006, it is one of the main references among the community, providing coordinates, proper motion, type, chart, spectral and period references for all 1830 sources when available. In order to obtain the purest sample, we discarded from this catalogue the objects designated as ``NON-CV'', which are stars that have been previously identified as CVs but later confuted, and those with the extensions ``:'' and ``::'' because their classification is not conclusive.

The Catalog of Cataclysmic Binaries, Low-Mass X-Ray Binaries and Related Objects \citep{RKCat} which only contains objects with a measured period, is updated up to December 31st, 2015 and it provides coordinates, apparent magnitudes, orbital parameters, stellar parameters of the components and other characteristic properties for 1429 CVs. In this case uncertain values are followed by only one ``:'' and have been discarded as well. 

Both catalogues have been merged into a final sample of 1920 CVs, out of which 1187 are contained in the GDR2 footprint. The density studies of CVs in the HR-diagram 
were done using this full sample. For 839 of these systems, the orbital period is known, and for 1130 systems, the subtype is unambiguously known (see Tab.\ref{tab:subtypes_sample}).


\begin{table}
\scriptsize
	\centering
	\caption{\label{tab:subtypes_sample} Distribution of the CV sample utilised by subtype.}
	\begin{tabular}{lcccc} 
		\hline
CV subtype & Periods &  Main & \multicolumn{2}{c}{Centroid position in HRD}\\
& sample & sample & $G_{BP}-G_{RP}$ & $G_{abs}$\\
		\hline
		Novalike & 76 & 119 & 0.37 & 5.63 \\
		Dwarf Novae & 484 & 688 & 0.64 & 9.49 \\
		Old Novae & 77 & 119 & 0.79 & 5.58 \\
		Polar & 75 & 135 & 0.83 & 9.67 \\
		Intermediate Polar & 51 & 69 & 0.59 & 5.61 \\
		\hline
		Total Sample & 839 & 1130 & &\\
		\hline
	\end{tabular}
\end{table}

\section{CVs in the HR-diagram}\label{sec:HRdiagram}


\subsection{The impact of the orbital period}

\begin{figure*}
    \centering
	    \begin{tabular}{c c}
	        \includegraphics[width=0.45\textwidth]{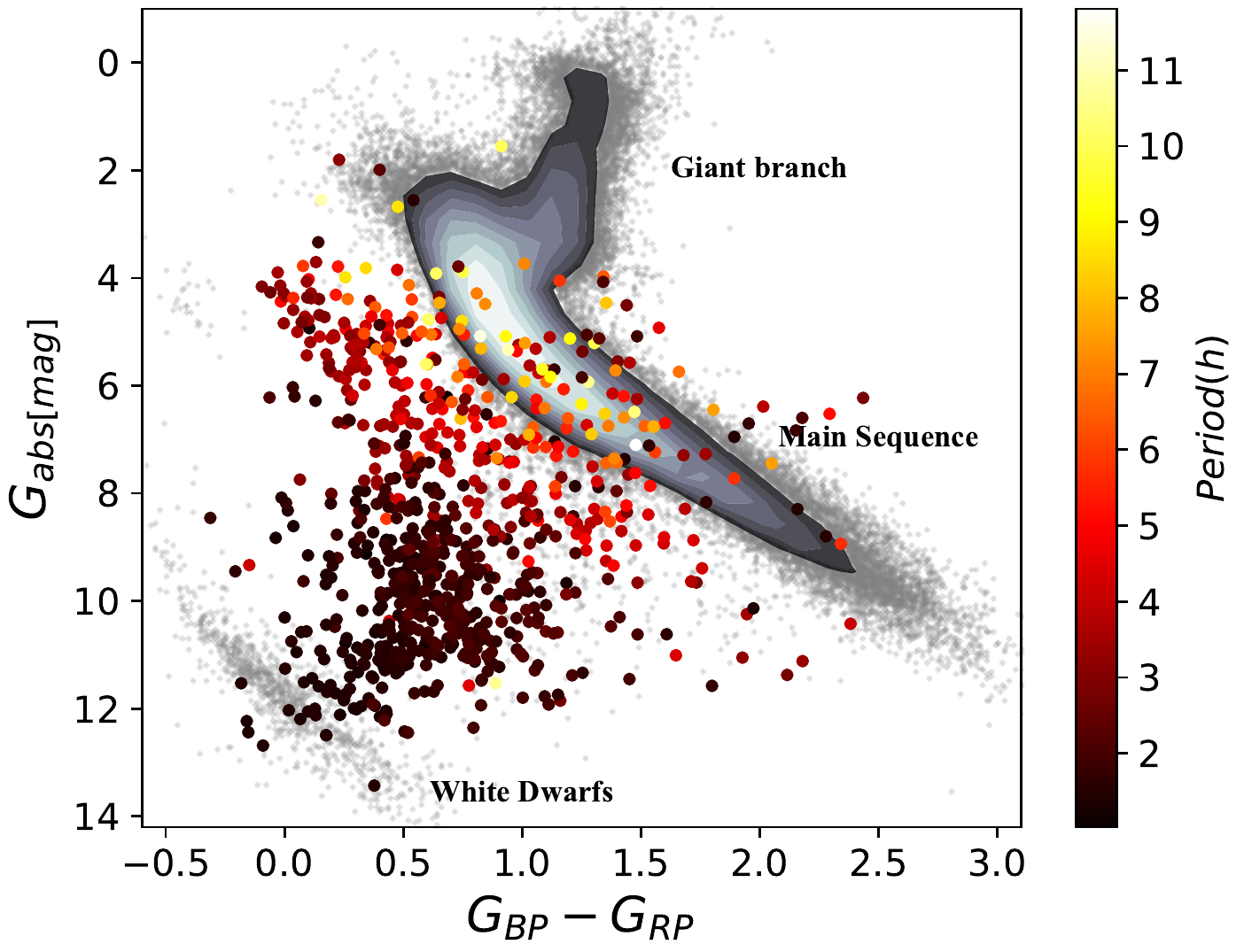}&	\includegraphics[width=0.42\textwidth]{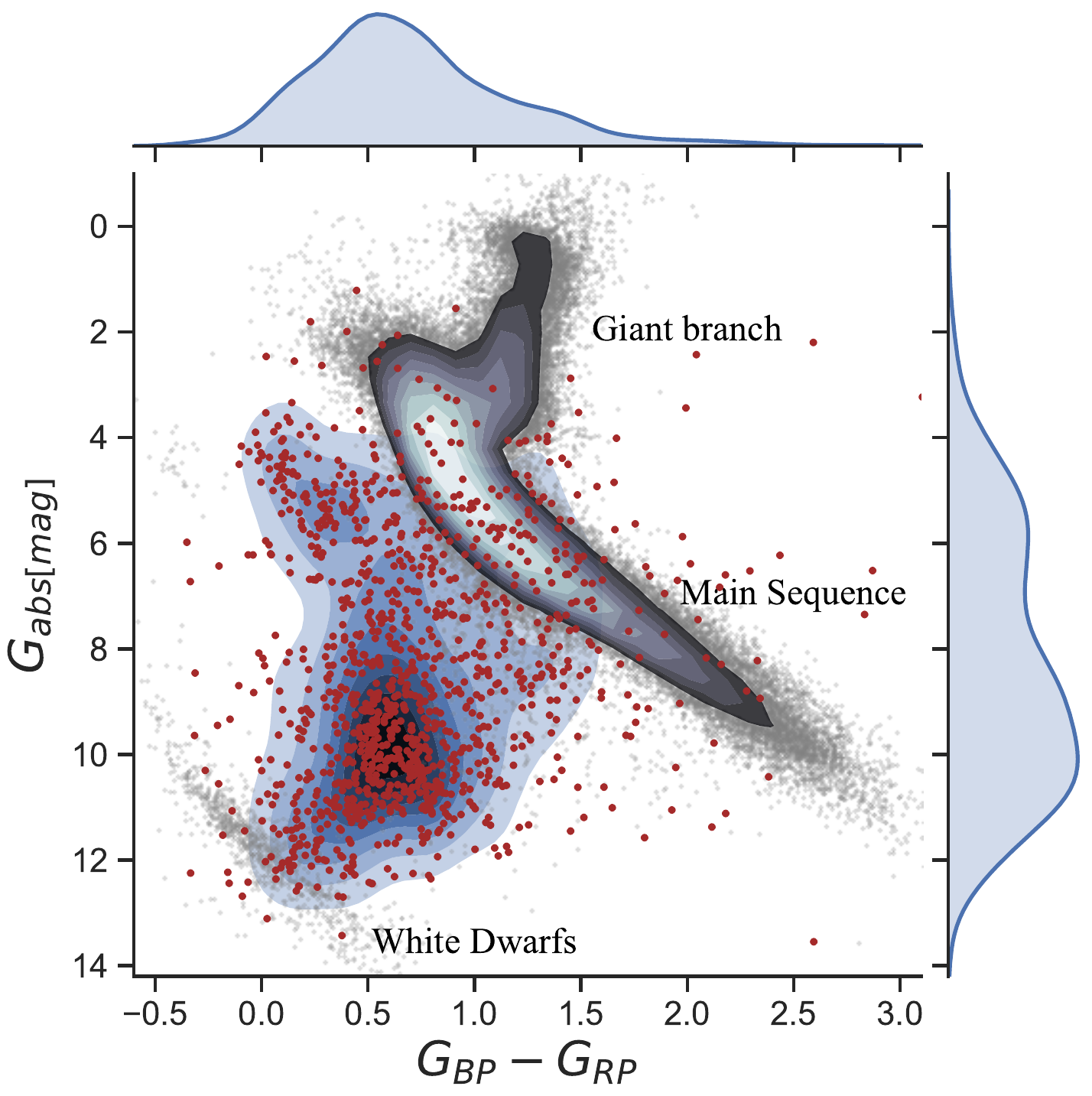}	
	    \end{tabular}
    \caption{In grey, all stars from {\it Gaia}'s 2nd data release are plotted in the HR-diagrams. On the left side, the CVs period distribution, the CVs from our sample with known orbital periods (see Section \ref{sec:CVsample}) are plotted in larger dots. The colour of each dot refers to the orbital period as given in the bar at the right of the panel. CVs with larger periods lie close to the main sequence path getting shorter while approaching the white dwarfs area. On the right, the density distribution of our whole sample of CVs (brown dots), surfaces with different tones of blue represent areas of equal density. On the x- and y-axis the marginal distributions are shown. CVs lie on average between the MS path and the WDs with a high density area peaking at $G_{BP}-G_{RP} \sim 0.56$ and $G_{abs} \sim 10.15$. Such area corresponds to the overpopulation below the period gap as reflected in the left panel by black dots, CVs with orbital period below 2h.
    \label{fig:HRper_densityMap}}
\end{figure*}

Left panel of Fig. \ref{fig:HRper_densityMap} displays the CV locus in the HR-diagram of all CVs for which an orbital period is known (839 systems).  The orbital period of each system is represented by the colour of the symbol as defined in the auxiliary axis. The CVs lie on average between the main sequence stars and the WDs. 
A clear trend is seen on their position with the orbital period: CVs with longer periods fall close to the main sequence path, while, as the orbital period decreases, they approach the WDs region.
This behaviour can be understood from the contribution of the secondary star. On average, a Roche-lobe filling secondary star is larger and brighter for longer orbital periods, while the WD does not change much during the secular CV evolution. Hence, the contribution of the secondary should be more dominant for longer orbital periods. Systems below the period gap, are instead dominated by their WD, as the secondary becomes only visible in the near infrared and does not contribute to the {\it Gaia} colour. The contribution of the accretion disc should change colour and magnitude depending on the sub-type and will be discussed in the next subsection.

The right panel of Fig. \ref{fig:HRper_densityMap} shows the locus of all CVs of our sample defined in Section \ref{sec:CVsample} within {\it Gaia's} HR-diagram. On the x- and y-axis, the respective projected density is plotted. A high density area is well distinguishable at $G_{BP}-G_{RP} \sim 0.56$ and $G_{abs} \sim 10.15$ (values obtained from the mode of the marginal distributions) which corresponds to the population below the period gap. 


\subsection{The locus depending on the subtype}
\begin{figure*}
\center
	\begin{tabular}{c c c}
	\includegraphics[width=0.3\textwidth]{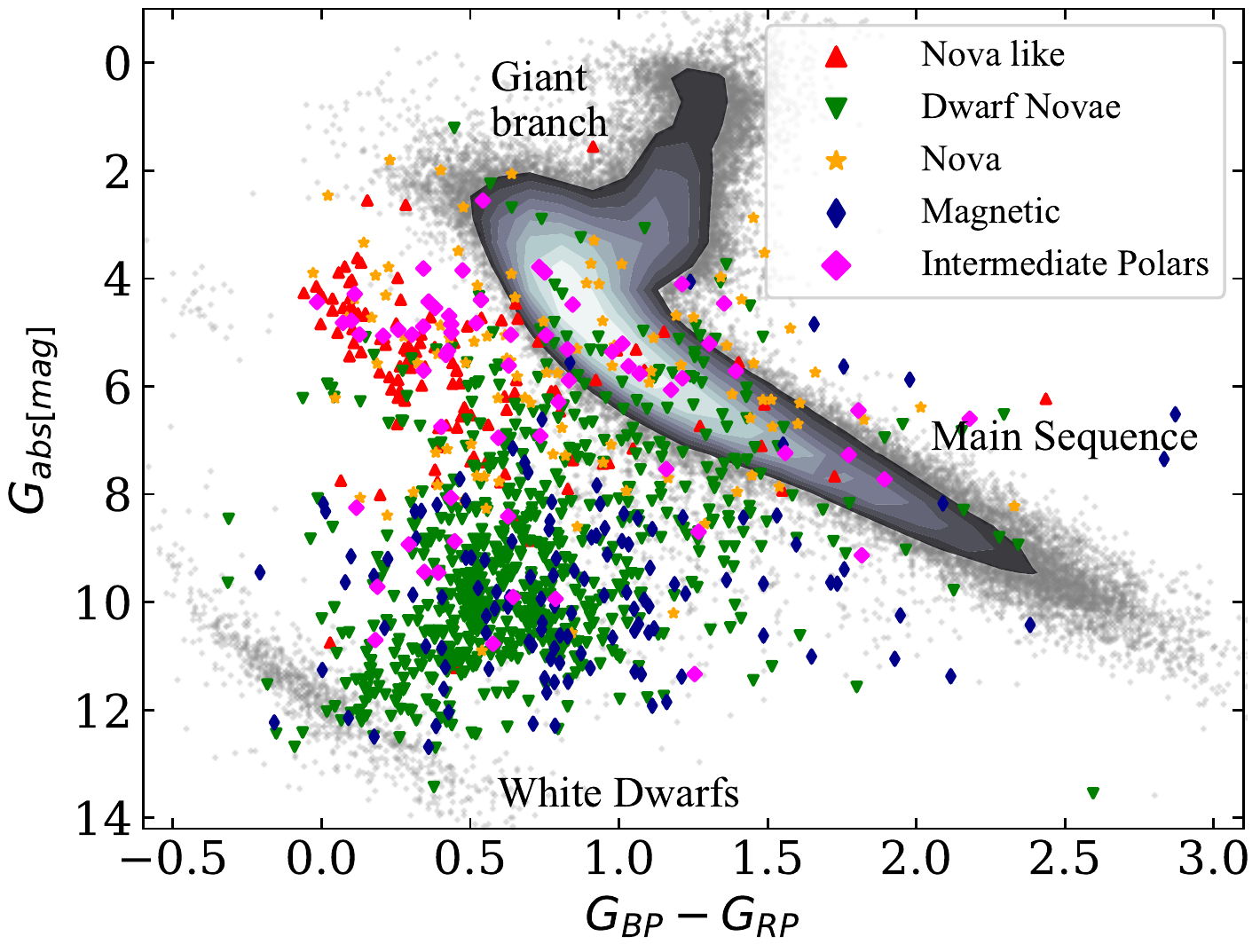}	&
	\includegraphics[width=0.3\textwidth]{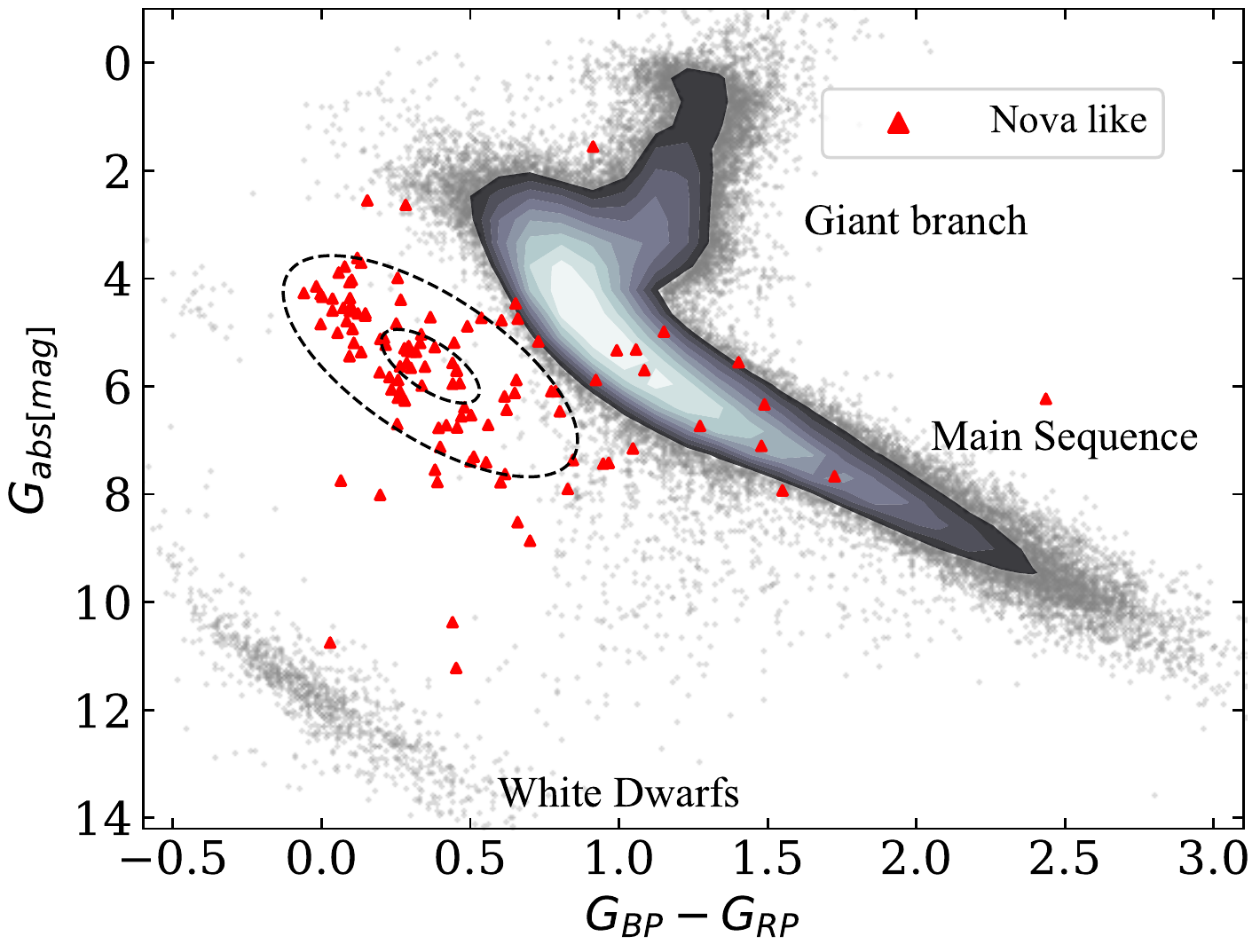}	&
	\includegraphics[width=0.3\textwidth]{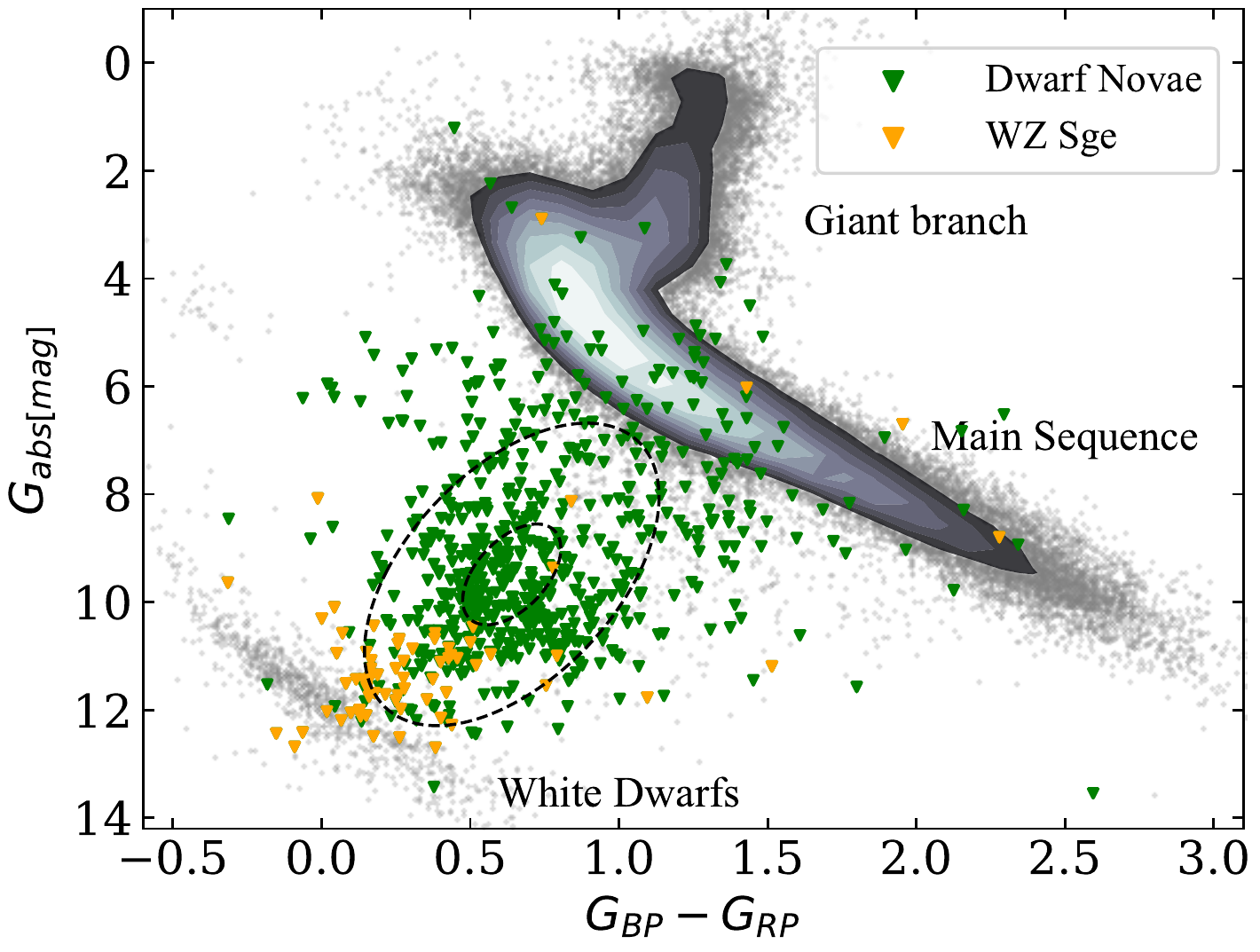}	\\
	\includegraphics[width=0.3\textwidth]{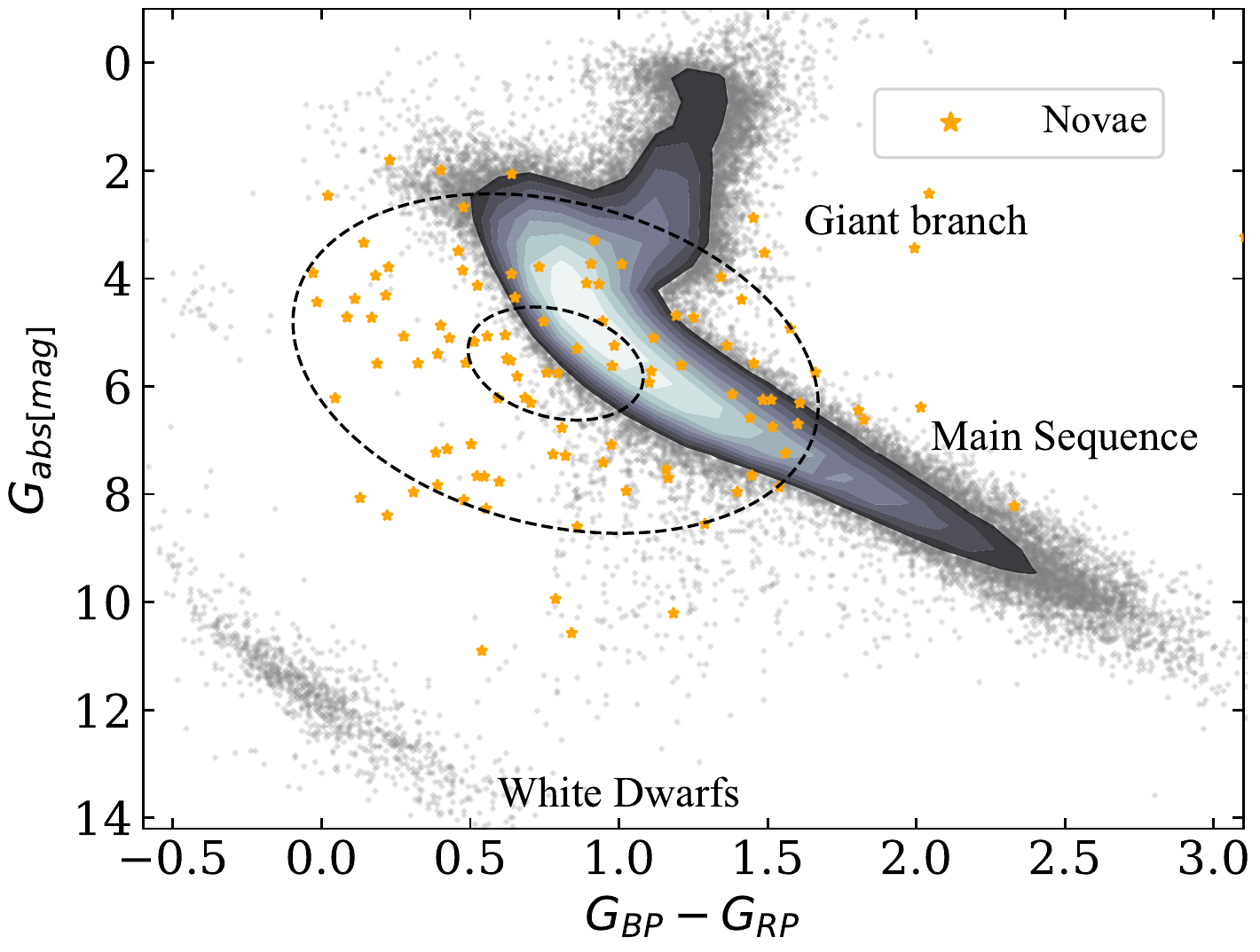}	&
	\includegraphics[width=0.3\textwidth]{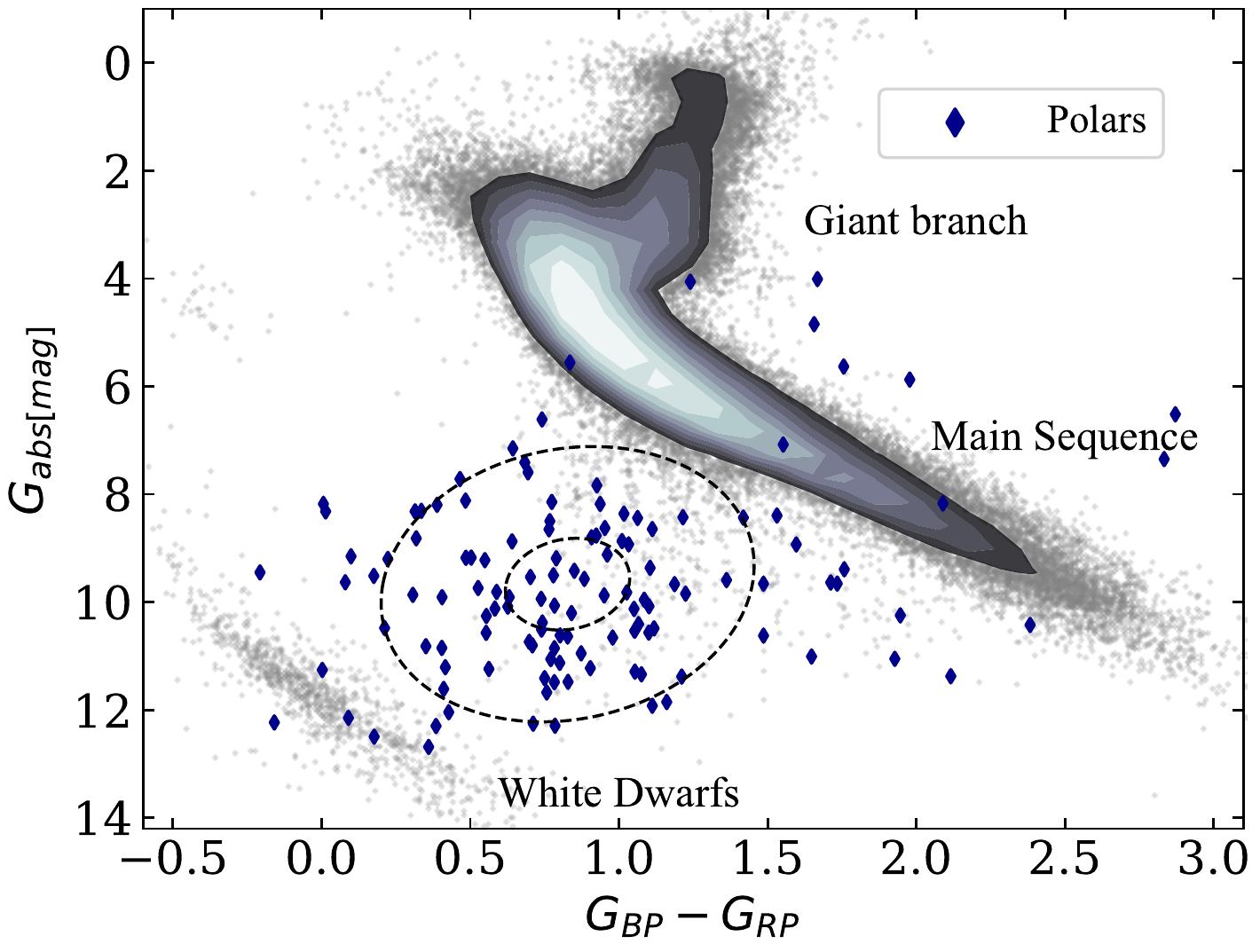}	&
	\includegraphics[width=0.3\textwidth]{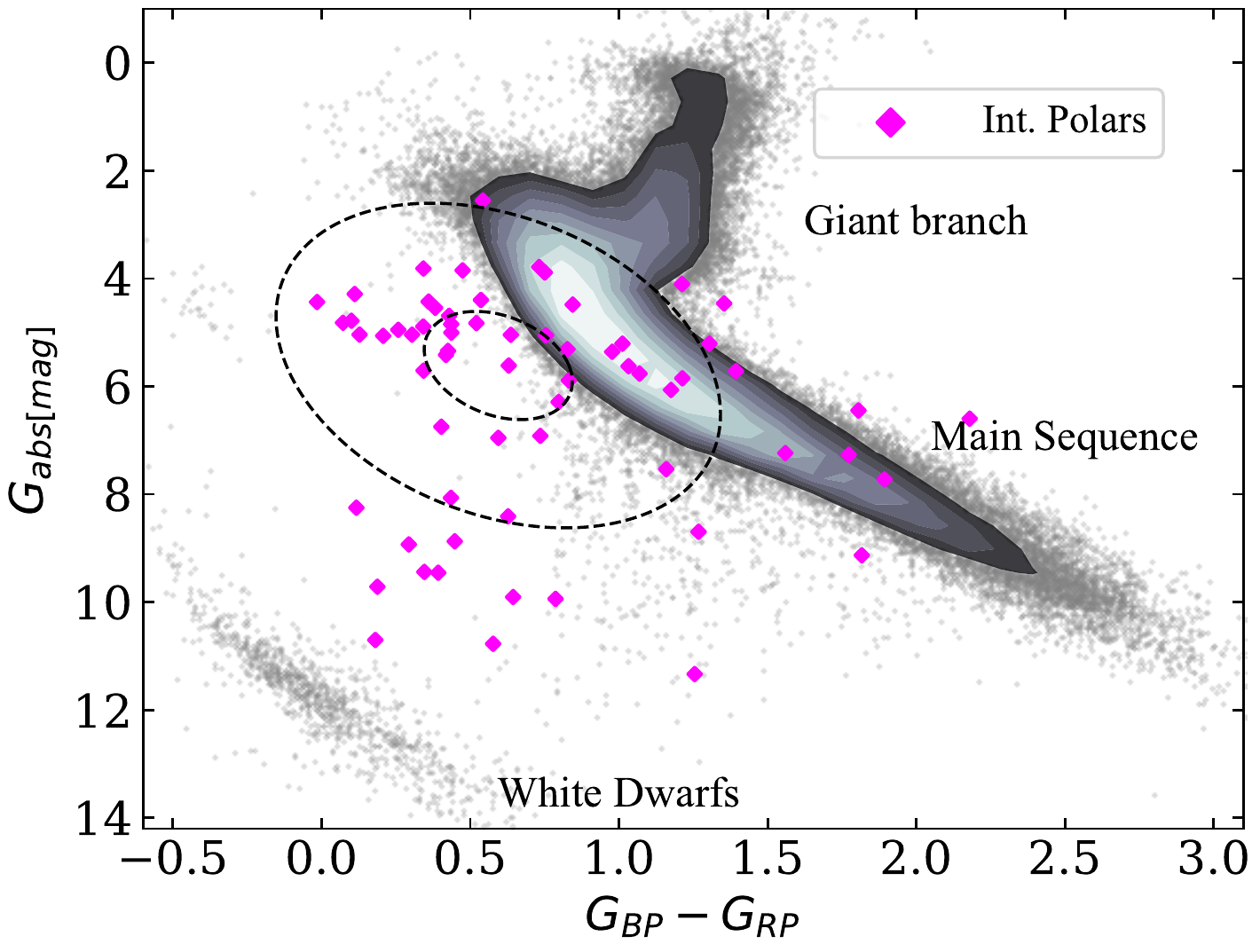}	\\
	\includegraphics[width=0.3\textwidth]{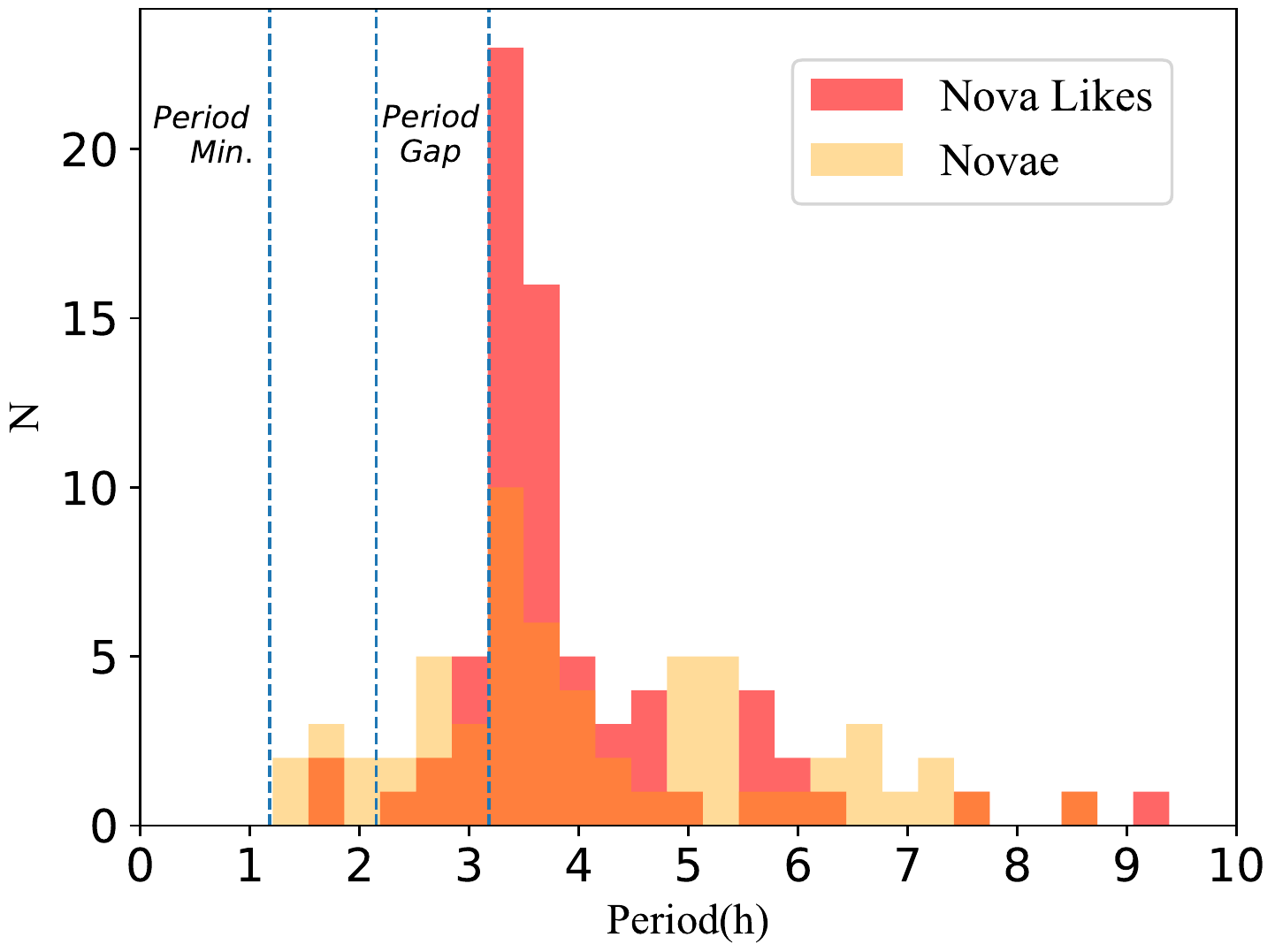}	&
	\includegraphics[width=0.3\textwidth]{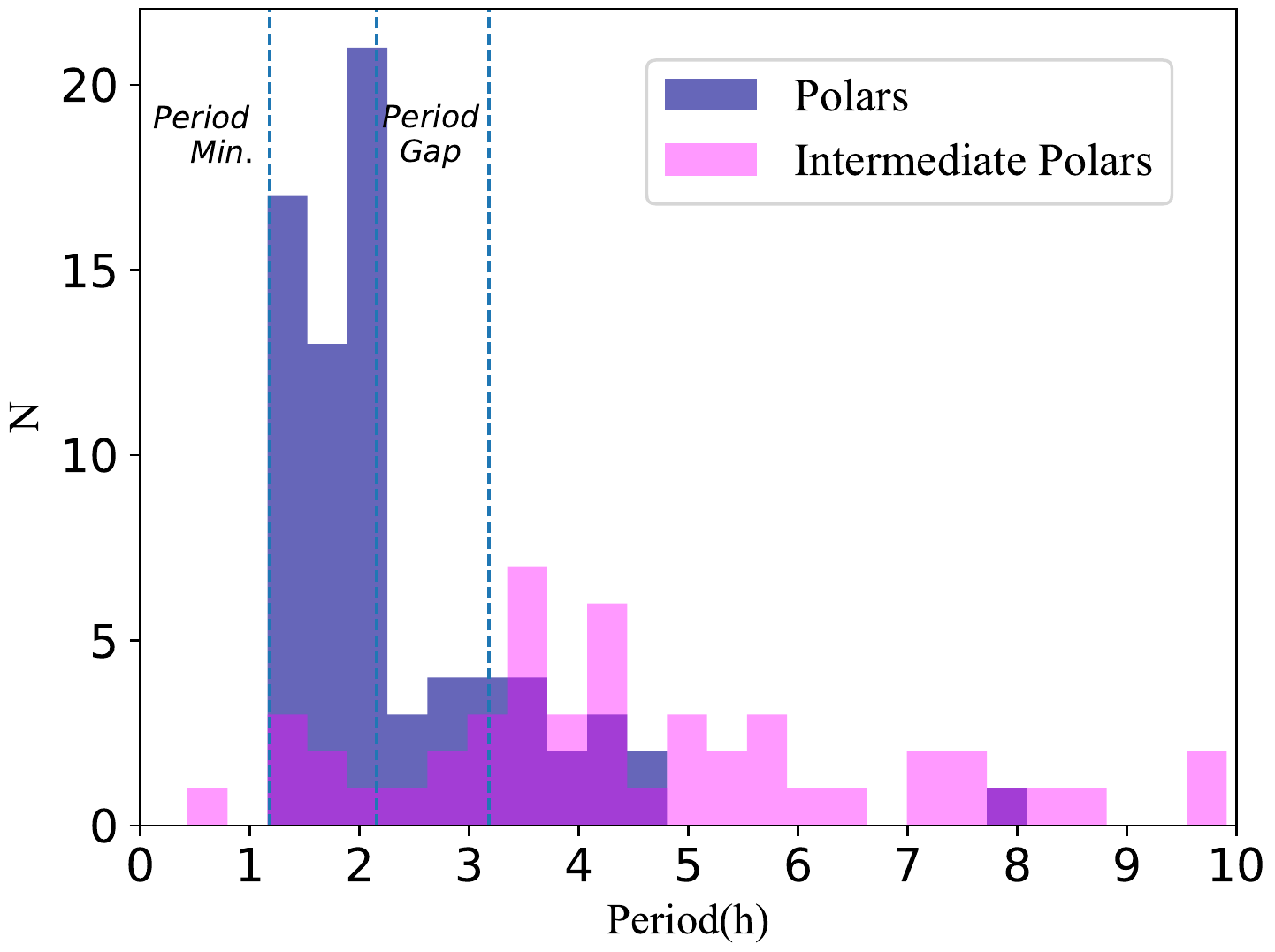}	&
	\includegraphics[width=0.3\textwidth]{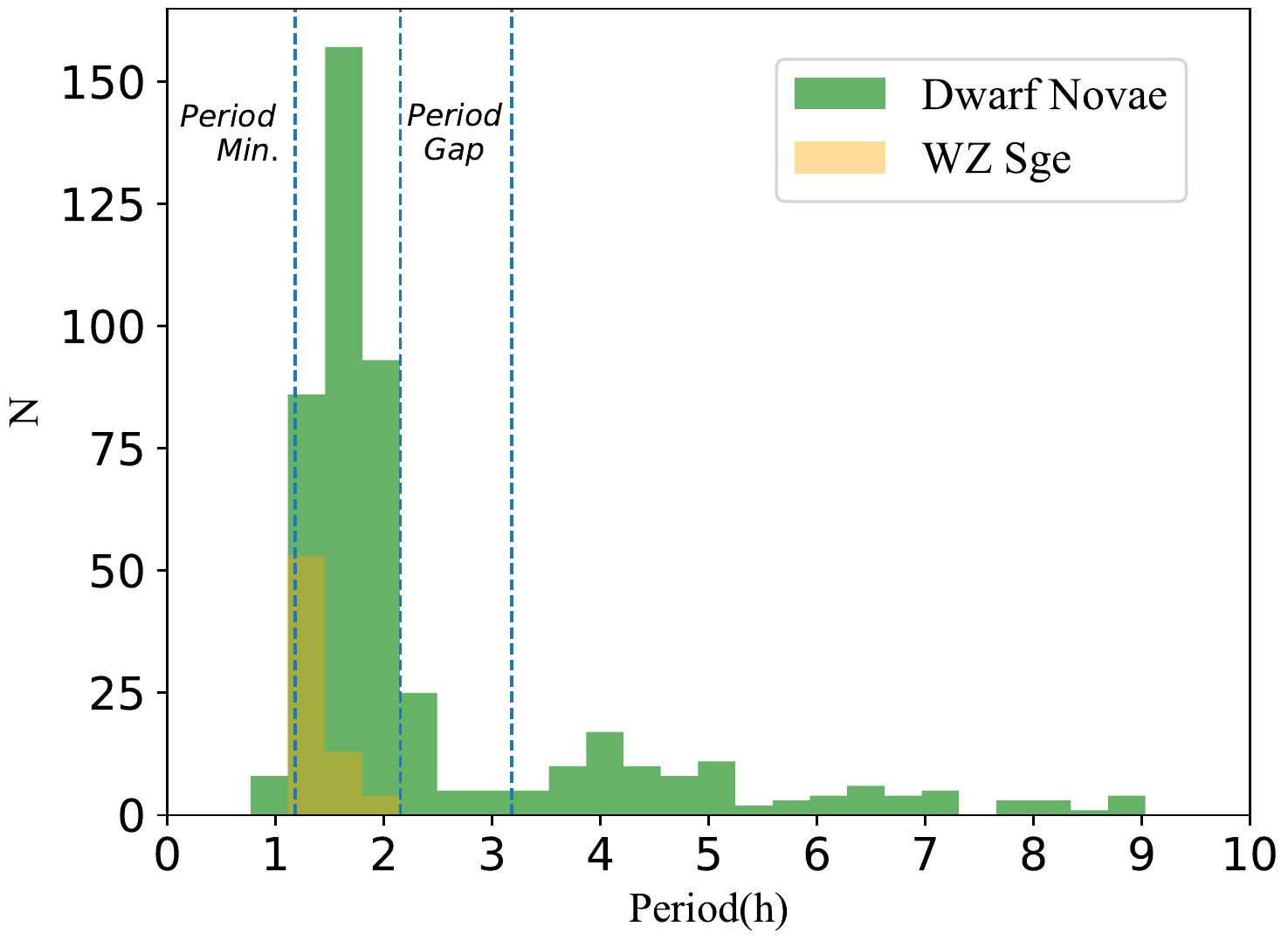}	
	\end{tabular}
\caption{The distribution of CV subtypes in the HR-diagram. Top-left panel shows all subtypes together, after that every subtype separately. The dashed ellipses represent 1 and 3 $\sigma$ of each subtype bivariate Gaussian distribution. The sample utilized here is composed by all CVs in the Ritter\&Kolb and Downes catalogues (see Section \ref{sec:CVsample}) whose subtype is unambiguously known and are included in the {\it Gaia} footprint; 119 Nova-likes, 688 Dwarf Novae, 119 Novae, 135 Polars and 69 Intermediate Polars. On the bottom, the period histograms for each subtype.  \label{fig:HR_per_types}}
\end{figure*}

Figure \ref{fig:HR_per_types} exhibits the distribution of every CV subtype on the HR-diagram. Bivariate Gaussian distributions are computed for 1 and 3 $\sigma$ given by

\begin{equation}
p(x, y | \mu_x, \mu_y, \sigma_x, \sigma_y, \sigma_{xy}) = \frac{1}{2 \pi \sigma_x \sigma_y \sqrt{1-\rho^2}}exp \left(\frac{-z^2}{2(1-\rho^2)}\right),
\label{eq:bivariate1}
\end{equation}

where

\begin{equation}
z^2 = \frac{(x-\mu_x)^2}{\sigma_x^2}+\frac{(y-\mu_y)^2}{\sigma_y^2}-2\rho \frac{(x-\mu_x)(y-\mu_y)}{\sigma_x \sigma_y},      
\label{eq:bivariate2}
\end{equation}

\begin{equation}
\rho = \frac{\sigma_{xy}}{\sigma_x \sigma_y},
\label{eq:bivariate3}
\end{equation}

using the median instead of the mean and the interquartile range to estimate variances in order to avoid the impact of outliers. The results are given in Table \ref{tab:subtypes_sample}.

Nova-likes are dominated by a high mass-transfer accretion disc, that usually 
overshines
the WD and the secondary star at optical and even infrared wavelengths. Their colour and final absolute magnitude mainly depends on the inclination with respect to the line of sight. In the HR-diagram, they concentrate around $G_{abs}=5.63$ and $G_{BP}-G_{RP}=0.37$, i.e. on the blue and bright corner of all CVs. A similar locus but with a much higher scatter is occupied by the old novae and by intermediate polars. This can be explained by the eclectic composition of these two sub-groups which also contain a large fraction of novalike stars. 

In contrast, polars which do not accrete mass through a disc, are much fainter and their colour and magnitude will depend on the nature of the secondary. In the HR-diagram they scatter around $G_{abs}=9.67$ and $G_{BP}-G_{RP}=0.83$ 
representing the reddest and faintest of all the CV subgroups.

Dwarf novae (DNe) occupy the whole region between MS stars and WDs with the centroid being at $G_{abs}=9.49$ and $G_{BP}-G_{RP}=0.64$. Since the secondary star in these systems can be anything from an early K-type star down to a brown dwarf, the range in colours and magnitude is not surprising. In addition, these sources are characterised by undergoing 
regular outbursts increasing their brightness and blueness. 
As discussed in Section \ref{sec:gaia}, the given magnitude is a weighted mean of several epochs and thus also increases the spread of this distribution. A detailed study of the DNe locus depending on their subtype and outburst state can be done following the next {\it Gaia} release when individual measurements and epochs become available.


WZ Sge-type objects deserve a separate mention, a class of DNe characterised by great outburst amplitudes, slow declines and long intervals between outbursts compared with ordinary DNe. These kind of systems have been considered to be period bouncer candidates \citep{Patterson11}, some of them extensively investigated in this regard (QZ Lib, \citealt{Pala18}; J122221 \citealt{vitaly17} and \citealt{kato13b}; J184228 \citealt{kato13b}; J075418 and J230425 \citealt{Nakata14}). We have plotted a sample of 71 of such systems in the upper-right panel of Fig. \ref{fig:HR_per_types}, along with the rest of DNe, and it can be seen that they concentrate near the WDs area. This is consistent with them being period bouncers or similar systems, as these are the CVs with the lowest mass transfer and faintest secondary stars. The disc is only visible in some emission lines, the secondary does not contribute to the optical range at all.

\subsection{Detached CVs}

\begin{figure}
	\includegraphics[width=0.425\textwidth]{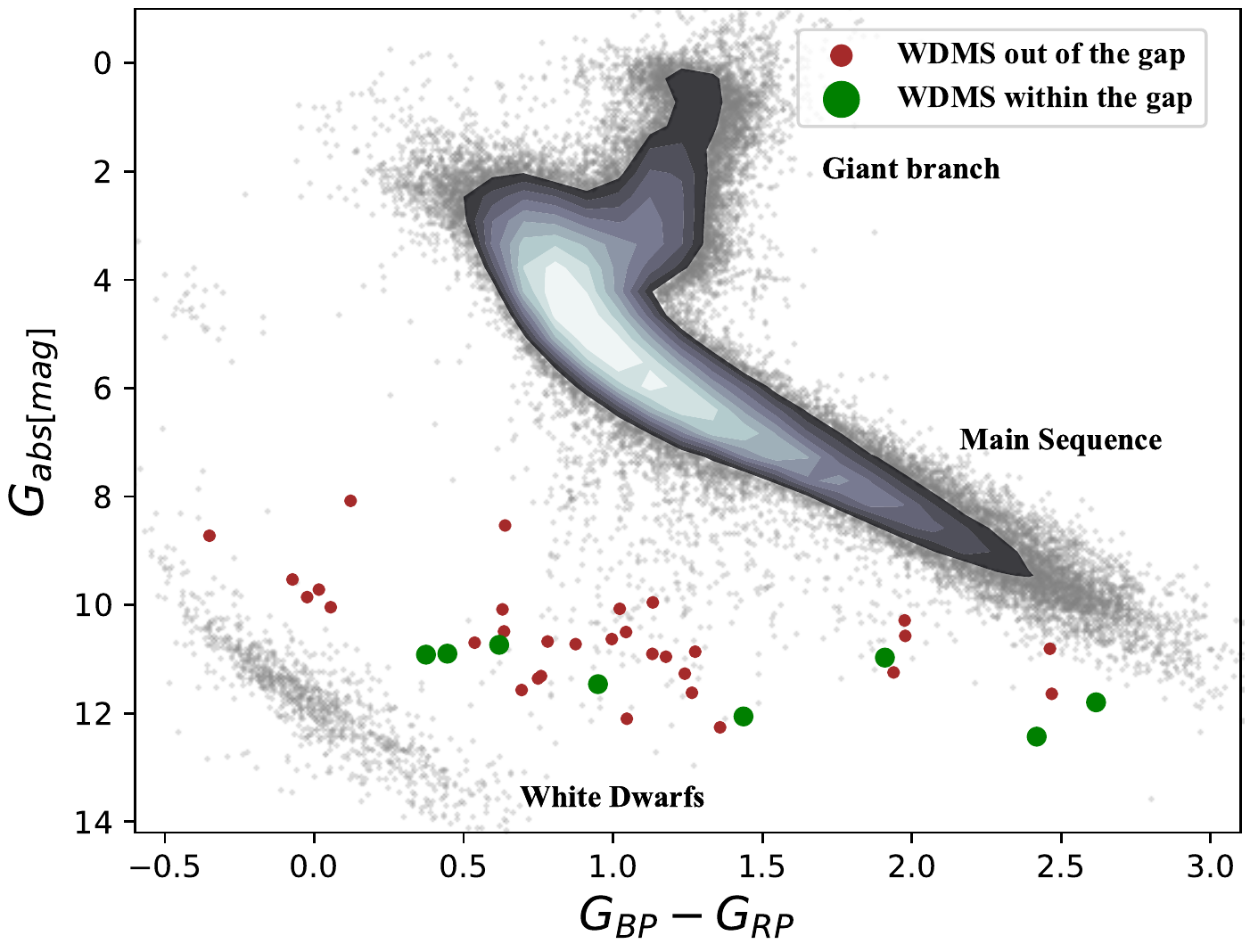}
\caption{\label{fig:dCVs} Systems comprised by a WD and an M4-6. The figure contains the 39 objects from the sample 
by \citep{Zorotovic16} present in {\it Gaia} (see text). In green are those within the period gap and in brown the rest. All of them follow a trend but the objects within the gap are on average fainter. 
}
\end{figure}

Another question we can address is finding the locus occupied by the so-called detached cataclysmic variables (dCVs) crossing the orbital period gap. A first approach could be made by finding the area with boundaries in 2 and 3h in the left panel of Fig. \ref{fig:HRper_densityMap} using 
regression techniques. However, since dCVs no longer 
contain an accretion disc, they should appear fainter than regular CVs of the same period. 

Due to the continuous mass loss, the donor is being driven out of equilibrium and secondaries in CVs just above the period gap are bloated up to 30\% with respect to regular MS stars \citep{kniggeetal11-1}. When the mass transfer stops, the secondary shrinks towards its thermal equilibrium radius to nearly its equivalent for MS stars \citep{stevehowell-01} and hence we expect secondaries in dCVs to be comparable to single MS stars of the same type.

Since the mass transfer ceases, the mass and spectral type of the donor star stays constant during the interval in which the binary is detached. In regular CVs this happens at $M_{sec} = 0.2 \pm 0.02$ $M\odot$ \citep{knigge06-1} and spectral type $\sim M6$ \citep{rebassaetal2007}, though variations occur depending on the moment in which the CV started the mass transfer and the time passed as CV until the secondary becomes fully convective. In the extreme case of a CV starting the mass transfer within the period gap range, the donor star type will be that of a fully convective isolated M star, which, according to \cite{Chabrier-97}, occurs at $M_{sec} \sim0.35$ $M\odot$ and spectral type M4. We thus assume that the secondary of dCVs is in the range M4-M6.

So far the only observational evidence for the existence of dCVs come from \cite{Zorotovic16}, who show that the orbital period distribution of detached close binaries consisting of a WD and an M4-M6 secondary star cannot be produced by Post Common Envelope Binaries alone, but a contribution of dCVs is needed to explain the peak between 2 and 3h. They also show that the systems inside this peak have a higher average mass than would be expected for normal WDMS systems. Still, with only 52 such systems known in total (WDMS systems with secondary spectral types in the range M4-6 and orbital periods below 10h) and 12 between 2 and 3h, the significance is not very high.


We distinguish two groups, the sources with orbital periods corresponding to those of the period gap (2-3h) and therefore, more likely to be dCVs, and the rest with periods outside this range. In Fig. \ref{fig:dCVs} they are plotted in the HR-diagram, the former appear 
fainter compared to the latter. This can be explained by the higher WD masses in CVs, and consequently in dCVs, than in PCEBs, making them smaller in size and surface and contributing in a lesser extent on the brightness 
of the whole system. 

\section{Summary}\label{sec:Summary}
We have analyzed the evolutionary cycle of CVs from a statistical perspective using {\it Gaia} DR2 data in conjunction with the HR-diagram tool. We have reported the discovery of a trend of the period and mass accretion with colour and absolute magnitude. We have also investigated their density distribution as a whole population, peaking at $G_{BP}-G_{RP} \sim 0.56$ and $G_{abs} \sim 10.15$, and the contribution of the main CV subtypes to this regard, highlighting the location of WZ Sge systems, which are period bouncer candidates. Finally, we have identified the location and a trend among systems comprised of a WD and secondary in the range M4-M6, which correspond with dCVs, CVs going through the orbital period gap.

\section*{Acknowledgements}
This research has made use of the VizieR catalogue access tool, CDS, Strasbourg, France (DOI: 10.26093/cds/vizier). The original description of the VizieR service was published in 2000, A\&AS 143, 23.




\bibliographystyle{mnras}
\bibliography{aabib} 




\appendix




\bsp	
\label{lastpage}
\end{document}